\documentstyle[12pt,fleqn]{article}
\catcode`\@=11
\begin{document} \openup6pt

\title{\Large{Naked Singularities in the Charged Vaidya-deSitter Spacetime}}
\author{A.~Beesham\thanks{Author to whom all correspondence should be
directed.  Tel: +27-35-7933911x2475; Fax: +27-35-7933079/7933735;
email: abeesham@pan.uzulu.ac.za} \and
S.G.~Ghosh\thanks
{Permanent Address: Department of Mathematics, Science College, Congress
Nagar, Nagpur - 440 012; email: sgghosh@pan.uzulu.ac.za} \\
Department of Applied Mathematics, University of Zululand, \\
Private Bag X1001, Kwa-Dlangezwa 3886, South Africa}

\vspace{2in}
\date{}
\maketitle



\begin{abstract}
We study the occurrence of naked singularities in the spherically symmetric
collapse of a charged null fluid in an expanding deSitter background -
a piece of charged Vaidya-deSitter spacetime.  The necessary conditions
for the formation of a naked singularity are found.  The
results for the uncharged solutions can be recovered from our analysis.
\end{abstract}



\section{Introduction}
In recent years there has been much interest in the problem of gravitational
collapse in general relativity (see \cite{r1} for recent reviews).  The end
state of gravitational collapse of a sufficiently massive body is a
gravitational singularity.  The conjecture that such a singularity from a
regular initial surface must always be hidden behind an event horizon
is called the cosmic censorship hypothesis (CCH) and was proposed
by Penrose \cite{rp}.  The weak CCH allows for the occurrence of locally
naked singularities but not globally naked ones, whereas the strong CCH
does not allow either.  This conjecture remains as the central open issue in
general relativity.  Hence, examples showing the existence of naked
singularities (which appear to violate the conjecture) remain important
and may be valuable when one attempts to formulate the notion of the CCH in
concrete mathematical terms.

In recent years significant attention has been given
to self similar spcetimes in the context of naked singularities \cite{ps1}.
However, self similarity is a strong geometric condition on the spacetime
and thus
gives rise to the possiblity that the naked singularity could be the result of
a geometric condition rather than the matter content of the spacetime.
So, it is useful to construct examples
which are not self similar and develop naked singularities.
The Vaidya solution \cite{pc} is most commonly used as a testing ground for
various violations of the CCH, e.g., the Vaidya solution representing
null dust collapse exhibits naked singularities \cite{r6}.  It was recently
shown \cite{wm} by means of an example that whether the space time is
asymptotically
flat or not does not make any difference to the occurrence of a locally naked
singularity.

The ingoing charged Vaidya solution \cite{bv} represents a
radial
infall of a stream of charged photons.  Lake and Zannias \cite{lz}, under the
assumption of homothecity, found that naked singularities can be formed.  The
usefulness of the model is that rich structure is exhibited.  In this context,
it is worthwhile to examine gravitational collapse of charged
radiation shells in an expanding deSitter background with reference to the
occurrence of naked singularities and the CCH.  The metric for this purpose
is already known \cite{ww}.
The organisation of the paper is as follows: In Section 2 we introduce the
charged Vaidya deSitter spacetime and note that the weak energy condition is
satisfied.  Section 3 is devoted in finding the analytical solution and
condition
for the existence of a naked singularity.  The paper ends with the discussion
in Section 4.

\section{Charged Vaidya-deSitter Spacetime}
The charged Vaidya-deSitter metric in $(v,r, \theta, \phi)$ coordinates is
\cite{ww}

\begin{equation}
ds^2 = - (1 -  \frac{2 m(v)}{r} + \frac{e^2(v)}{r^2} - \Lambda
\frac{r^2}{3}) dv^2 + 2 dv dr + r^2 d \Omega^2  \label{eq:me}
\end{equation}
where $ d\Omega^2 = d \theta^2+ sin^2 \theta d \phi^2$, $v$ represents
advanced Eddington time, in which $r$  is decreasing towards the future
along a ray $v=const.$, the two arbitrary functions $m(v)$ and $e(v)$
(which are resticted only by the energy conditions), represent, respectively,
the mass and electric charge at advanced time $v$, and $\Lambda$ is the
cosmological constant.  This metric (\ref{eq:me})
represents a solution to the Einstein
 equations for a collapsing charged null fluid in an expanding deSitter
background.

The energy momentum tensor can be written in the form
\begin{equation}
T_{ab} = T^{(n)}_{ab} + T^{(m)}_{ab} \label{eq:te}
\end{equation}
where
\begin{equation}
 T^{(n)}_{ab} = \frac{1}{4 \pi r^3} \left[ r \dot{m}(v) - e(v) \dot{e}(v)
\right] k_{a}k_{b} \label{eq:tn}
\end{equation}
with the null vector $k_{a}$ satisfying
$k_{a} = - \delta_{a}^{v} \; and \; k_{a}k^{a} = 0$,
$T_{ab}^{(m)}$ is related to the electromagnetic tensor $F_{ab}$:
\begin{equation}
T^{(m)}_{ab} = \frac{1}{4 \pi} \left( F_{ac} F_{b}^{c} - \frac{1}{4}
g_{ab} F_{cd} F^{cd} \right) \label{eq:tm}
\end{equation}
which satisfies Maxwell's field equations
\begin{equation}
F_{[a b;c]} = 0 \, and \, F_{a b;c} g^{bc} = 4 \pi J_{a} \label{eq:mf}
\end{equation}
where $J_{a}$ is the four-current vector.

Clearly, for the weak energy
condition to be satisfied we require the bracketed quantity in eq. (3) to be
non negative.  We note that the stress tensor in general may not obey the
weak energy condition.  In particular, if $dm/de > 0$ then there always
exists a
critical radius $r_{c} = e \dot{e}/ \dot{m}$ such that when $r < r_{c}$
 the weak energy condition is always violated.  However, in realistic
situtions, the particle cannot get into the region $r < r_{c}$ because of
the Lorentz force and so the energy condition is still preserved \cite{ww,oa}.

The Kretschmann scalar for the metric (\ref{eq:me}) reduces to
\begin{equation}
K = \frac{48}{r^6} \left[m^2(v) - \frac{2}{r} e^2(v)m(v) + \frac{7}{6}
 \frac{e^4(v)}{r^2}  \right] + \frac{8}{3} \Lambda^2   \label{eq:ks}
\end{equation}
So the Kretschmann scalar diverges along $r = 0$, establishing that
metric (\ref{eq:me}) is scalar polynomial singular.

\section{The Existence and Nature of Naked Singularities}
The physical sitution is that of a radial influx of charged null fluid
in an initially empty region of the deSitter universe.  The first shell
arrives at $r=0$ at time $v=0$ and the final at $v=T$.
A central singularity of growing mass is
developed at $r=0$.
  For $ v < 0$ we have $m(v)\;=\;e(v)\;=\;0$, i.e., an empty deSitter metric,
 and for $ v > T$,
$\dot{m}(v)\;=\;\dot{e}(v)\;=\;0$, $m(v)\; and \;e^2(v) $ are positive
definite.  The metric for $v=0$ to $v=T$ is  charged
Vaidya-deSitter, and for $v>T$ we have the Reissner
Nordstr$\ddot{o}$m deSitter solution.

In order to get an analytical solution, we choose $m(v) \propto v
\; and \;e^2(v) \propto v^2 $ \cite{me}. To be specific we take
\begin{equation}
m(v) = \left\{ \begin{array}{ll}
 0,   & \mbox{$ v < 0$}, \\
 \lambda v (\lambda>0) & \mbox{$0 \leq v \leq T$}, \\
 m_{0}(>0)  & \mbox{$v >  T$}.
  \end{array}
 \right.    \label{eq:mv}
\end{equation}
and
\begin{equation}
e^2(v) = \left\{ \begin{array}{ll}
 0,   & \mbox{$v < 0$}, \\
 \mu^2 v^2 (\mu^2>0) & \mbox{$0 \leq v \leq T$}, \\
 e_{0}^2 (>0)  & \mbox{$v >  T$}.
  \end{array}
 \right.    \label{eq:ev}
\end{equation}
When $\Lambda = 0$, the spacetime is self similar, admitting a
homothetic Killing vector.  However if $\Lambda \neq  0$, the basic
requirement of self similarity \cite{ss} is not met.  So, the line element
(\ref{eq:me})
does not admit any proper conformal Killing symmetries.

Radial ($ \theta$ and $ \phi \,=\,const$.) null
geodesics of the metric (1) must satisfy the null condition
\begin{equation}
\frac{dr}{dv} = \frac{1}{2} \left[1 -  \frac{2 m(v)}{r} + \frac{e^2(v)}{r^2}
- \Lambda \frac{r^2}{3} \right]  \label{eq:de1}
\end{equation}
which, upon using  eqs. (\ref{eq:mv}) and (\ref{eq:ev}) turns out to be
\begin{equation}
\frac{dr}{dv} = \frac{1}{2} \left[1 - 2 \lambda X + \mu^2 X^2
- \Lambda \frac{r^2}{3} \right]  \label{eq:de2}
\end{equation}
where $X \equiv v/r$ is the tangent to a possible outgoing geodesic.
Clearly, the above differential equation has a singularity at $r=0$, $v=0$.
The nature (a naked singularity or a black hole) of the collapsing solutions
can be characterised by the existence of radial null geodesics coming out from
the singularity.

In order to determine the nature of the limiting value of $X$ at $r=0$, $v=0$
on a singular geodesic, we let
\begin{equation}
X_{0} = \lim_{r \rightarrow 0 \; v\rightarrow 0} X =
\lim_{r\rightarrow 0 \; v\rightarrow 0} \frac{v}{r}     \label{eq:lm1}
\end{equation}
Using (\ref{eq:de2}) and L'H\^{o}pital's rule we get
\begin{equation}
X_{0} = \lim_{r\rightarrow 0 \; v\rightarrow 0} X =
\lim_{r\rightarrow 0 \; v\rightarrow 0} \frac{v}{r}=
\lim_{r\rightarrow 0 \; v\rightarrow 0} \frac{dv}{dr} =
\frac{2}{1 - 2 \lambda X_{0} + \mu^2 X_{0}^2}  \label{eq:lm2}
\end{equation}
which implies,
\begin{equation}
 \mu^2 X_{0}^3 - 2 \lambda X_{0}^2 + X_{0} - 2 = 0    \label{eq:ae}
\end{equation}
This algebraic equation governs the behaviour of the tangent vector near the
singular point.  Thus by studying the solution of this algebraic equation,
the nature of the singularity can be determined.  The central shell focussing
singularity is at least locally naked (for brevity we have addressed it as
naked throughout this paper), if eq. (\ref{eq:ae})
 admits one or more positive
real roots.  However, the locally naked singularities should not be treated
less seriously \cite{rp1}.
When there are no positive real roots to eq. (\ref{eq:ae}),
the central
singularity is not naked because in that case there are no out going future
directed null geodesics from the singularity.  Hence in the absence of
positive real roots, the collapse will always lead to a black hole.
The condition under which this locally naked singularity could be globally
naked is well discussed \cite{ps1} and we shall not discuss it here.
Thus, the
occurrence of positive real roots implies that the strong CCH is violated,
though
 not necessarily the weak CCH.  We now examine the condition for the
occurrence of a naked singularity.

We know that a cubic equation admits at least one real root.  Eq. (\ref{eq:ae})
 will have three real roots if
$\lambda^2 + 18 \lambda \mu^2 \geq 16 \lambda^3 + \mu^2 + 27 \mu^4$.
 Howevever, since $\lambda>0$ and $\mu^2 > 0$, eq. (\ref{eq:ae}) cannot have
any negative roots.  Hence eq. (\ref{eq:ae}) has at least one positive real
root.  It follows that the gravitational collapse of a charged null fluid
must lead to a naked singularity.

The Kretschmann scalar with the help of  eqs. (\ref{eq:mv}) and
(\ref{eq:ev}), takes the form
\begin{equation}
K = \frac{48}{r^4} \left( \lambda^2 X^2 - 2 \lambda \mu^2 X^3 + \frac{7}{6}
\mu^4 X^4 \right)+ \frac{8}{3} \Lambda^2 \label{eq:ks1}
\end{equation}
which diverges at the naked singularity and hence the singularity is a scalar
polynomial singularity.

Having seen that the naked singularity in our model is a scalar polynomial
singularity, we now turn our attention to the Weyl tensor.  It is
well known that the Weyl tensor vanishes at all points for any conformally
flat
space time.  The divergence of the Weyl tensor at the singularity could imply
 that such singularities are not associated with the matter distribution
and hence are  to be taken seriously (see \cite{bs1}, for more details).

Ths surviving components of the Weyl tensor are
\[
C_{0101} = \frac{2}{r^2} \left[ \frac{-m(v)}{r} + \frac{e^2(v)}{r^2} \right]
\]
\[
C_{0202} = \frac{m(v)}{r} \left[1 - \frac{2m(v)}{r} + \frac{3e^2(v)}{r^2}
- \frac{\Lambda}{3} r^2 \right] -
\frac{e^2(v)}{r^2} \left[1+
\frac{e^2(v)}{r^2} - \frac{\Lambda}{3} r^2 \right]
\]
\[
C_{0212} = \left[ \frac{-m(v)}{r} + \frac{e^2(v)}{r^2} \right]
\]
\begin{eqnarray}
C_{0303} = \frac{m(v)}{r} \left[1 - \frac{2m(v)}{r} + \frac{3e^2(v)}{r^2}
- \frac{\Lambda}{3} r^2 \right] sin^2 \theta -         \nonumber \\
\frac{e^2(v)}{r^2} \left[1+
\frac{e^2(v)}{r^2} - \frac{\Lambda}{3} r^2 \right] sin^2 \theta  \nonumber
\end{eqnarray}
\[
C_{0313} = \frac{2}{r^2} \left[ \frac{-m(v)}{r} + \frac{e^2(v)}{r^2}
\right] sin^2 \theta
\]
\[
C_{2323} = -2 \left [-r m(v) + e^2(v) \right] sin^2 \theta
\]
The Weyl scalar ($C = C_{abcd} C^{abcd}$, $C_{abcd}$ is the Weyl tensor),
upon inserting eqs. (\ref{eq:mv}) and (\ref{eq:ev}), reads
\begin{eqnarray}
C(v,r) = \frac{48}{r^6} \left[m^2(v) - \frac{2 m(v) e^2(v)}{r}
+ \frac{e^4(v)}{r^2} \right]   \nonumber \\
= \frac{48}{r^6} \left[ \lambda^2 v^2 - \frac{2 \lambda \mu^2 v^3}{r}
+ \frac{\mu^4 v^4}{r^2} \right]   \label{eq:ws1}
\end{eqnarray}
It can be noted that the Weyl scalar is zero in the deSitter region ($v < 0$).
  It is also zero at $r \neq 0$, $v=0$.  Eq. (\ref{eq:ws1}) can be written as
\begin{equation}
C(X,r) = \frac{48}{r^4} \left[ \lambda^2 X^2 - 2 \lambda \mu^2 X^3
+ \mu^4 X^4  \right]     \label{eq:ws2}
\end{equation}
Thus the Weyl scalar diverges at the naked singularity.  The effect of the
energy
momentum tensor on the geometry can be found by evaluating the Ricci scalar
($R =
R_{ab}^{ab}$, $R_{ab}$ the Ricci tensor) and comparing with the Weyl scalar
\cite{bs1}.
The Ricci scalar for the metric (\ref{eq:me}) is
\begin{equation}
R = 4 \left(\Lambda^2 + \frac{e^4(v)}{r^8} \right) =
4 \left(\Lambda^2 + \frac{\mu^4 X^4}{r^4} \right)   \label{eq:rs1}
\end{equation}
In contrast to the uncharged Vaidya space time \cite{bs1},
in this case, the Ricci scalar diverges at
the same rate as the Weyl scalar.  So the Weyl scalar does not dominate in our
case.

\section{Discussion}

We have shown the development of naked curvature singularities
in the charged Vaidya-deSitter spacetime.  In the limit $\mu
\rightarrow 0$, our results reduce to the those obtained
previously  \cite{wm} for the uncharged case.
Lake and Zannias \cite{lz} had
shown that naked singularities occur in the charged Vaidya
spacetime. We have shown that the asymptotic flatness of the spacetime does
not alter the result, i.e., naked singularities also occur if
$\Lambda \neq 0$. This agrees with the conclusion of Wagh and Maharaj
\cite{wm}, i.e., the occurrence of a naked singularity does not depend upon
whether the asymptotic spacetime is expanding or not.

Wagh and Maharaj \cite{wm} had shown that in the case of the uncharged
Vaidya-deSitter spacetime, a naked singularity occurs only if $\lambda
\leq 1/8$.  Most earlier work \cite{ps1} on naked singularities had shown
that they occur only for specific values of the parameters.  In contrast,
we have shown that for the charged case, a naked singularity always occurs,
irrespective of the value of $\lambda$.

We found that the Kretschmann scalar diverges in the limiting
approach to the singularity along radial null geodesics.  So,
the singularity can be considered as a physically significant
curvature singularity, and hence cannot be ignored.

The behaviour of the Ricci scalar is greatly affected by the
presence of charge.  It diverges at the same rate as the
Weyl scalar.  Therefore, one can say that the role played by
initial conditions in the deciding metric is equally important to
that of the energy momentum tensor whenever a naked singularity
occurs.  Further, the Weyl tensor is that part of the curvature of
spactime that is not locally detemined by matter, and hence the
divergence of the Weyl scalar at the naked singularity implies
that the singularity is not associated with the local matter
distribution and hence should be taken seriously.

In conclusion, we have obtained another example showing that the formation
of a naked singularity is not resticted to self similar spacetimes and
to an asymptotatically flat setting.
However, it remains to be shown whether or not the singularity
satisfies the strong curvature condition.

{\bf Acknowledgement:} One of the authors (SGG) would like to thank the
University of Zululand for hospitality, the NRF (South Africa) for financial
support and  Science College, Congress Nagar, Nagpur (India) for granting
leave.


\noindent

\begin{thebibliography}{99}
\bibitem{me} $m \propto v $ was introduced by A. Papapetrou (see Ref 5 above)
and subsequently used by many authors,  $e^2 \propto v^2$ has been examined
by Lake and Zannias (see Ref 7 above).
\bibitem{ss} A spherical symmetric space-time is self similar if
$ g_{tt} (ct, cr) = g_{tt}(t,r)$ and  $ g_{rr} (ct, cr) = g_{rr}(t,r)$ for
every $c>0$.  A self similar spacetime is characterised by the existence of
a homothetic Killing vector.
\end{thebibliography}
\end{document}